\documentclass[aps,prl,twocolumn,showpacs]{revtex4-1}
\usepackage{amsmath}
\usepackage{amsfonts}
\usepackage{amssymb}
\usepackage{epsfig}

\newcommand{\ra}{\rangle}
\newcommand{\la}{\langle}
\newcommand{\be}{\begin{equation}}
\newcommand{\ee}{\end{equation}}

\begin{document}

\title{Spontaneous Symmetry Breaking at the Fluctuating Level}

\author{Pablo I. Hurtado}
\author{Pedro L. Garrido}
\affiliation{
Departamento de Electromagnetismo y F\'{\i}sica de la Materia, and 
Instituto Carlos I de F\'{\i}sica Te\'orica y Computacional, Universidad de Granada, 
Granada 18071, Spain}

\date{\today}

\begin{abstract}
Phase transitions not allowed in equilibrium steady states may happen however at the fluctuating level.
We observe for the first time this striking and general phenomenon measuring current fluctuations in an isolated diffusive system.
While small fluctuations result from the sum of weakly-correlated local events, for currents above a 
critical threshold the system self-organizes into a coherent traveling wave which facilitates the current deviation by 
gathering energy in a localized packet, thus breaking translation invariance. 
This results in Gaussian statistics for small fluctuations but non-Gaussian tails above the critical current.
Our observations, which agree with predictions derived from hydrodynamic fluctuation theory, strongly suggest that 
rare events are generically associated with coherent, self-organized patterns which enhance their probability.
\end{abstract}

\pacs{05.40.-a, 11.30.Qc, 66.10.C-}

\maketitle

Fluctuations arise in most physical phenomena, and their study has proven once and again to 
be a fruitful endeavour. The first example is probably Einstein's determination of molecular scales on the 
basis of the fluctuating behavior of a mesoscopic particle inmersed in a fluid \cite{Einstein}, which opened 
the door to an experimental verification of the molecular hypothesis. Other examples range from 
the role of fluctuations to understand critical phenomena beyond mean field theories,
to the study of fluctuations of spacetime correlations in glasses and other disordered materials, 
which has revealed the universal existence of dynamic heterogeneities in these systems \cite{glass}.
In all cases the statistics of fluctuations encodes essential information to understand the physics of
the system of interest. Even further, fluctuations reflect the symmetries of the microscopic world at the macroscale. This is the 
case for instance of the Gallavotti-Cohen fluctuation theorem \cite{GC,K,LS} or the recently introduced isometric 
fluctuation relation \cite{IFR}, which express the subtle but enduring consequences of microscopic time-reversibility at the macroscopic level. 
Special attention is due to large fluctuations which, though rare, play a dominant role 
as they drastically affect the system behavior.

The study of fluctuating behavior provides an alternative way to derive thermodynamic potentials from which to calculate the
properties of a system, a path complementary to the usual ensemble approach. 
This can be extended to systems far from 
equilibrium \cite{Derrida,glass,GC,K,LS,IFR,Bertini,Bertini2,Bertini3,Jona}, where 
no bottom-up approach exists yet connecting microscopic dynamics with macroscopic properties.
The large-deviation function (LDF) controlling the fluctuations of the relevant macroscopic observables plays in nonequilibrium 
systems a role akin to the equilibrium free energy, and reflects the phenomenology typical of nonequilibrium physics 
(e.g. non-local behavior resulting in long-range correlations \cite{Derrida,Bertini}).
Hydrodynamic fluctuation theory (HFT), which studies dynamic fluctuations in diffusive media \cite{Bertini,Bertini2,Bertini3,Jona},
offers predictions for both the LDF and the optimal path in phase space responsible of a given fluctuation, which can be in general time-dependent \cite{Bertini2,Bertini3}. However, it has been shown that this optimal path is in fact time-independent in a broad regime \cite{BD,Pablo}.
This scenario eventually breaks down for large fluctuations via a dynamic phase transition \emph{at the fluctuating level} involving a symmetry breaking \cite{Bertini3,BD2}.

\begin{figure}[b]
\vspace{-0.6cm}
\centerline{
\includegraphics[width=8cm]{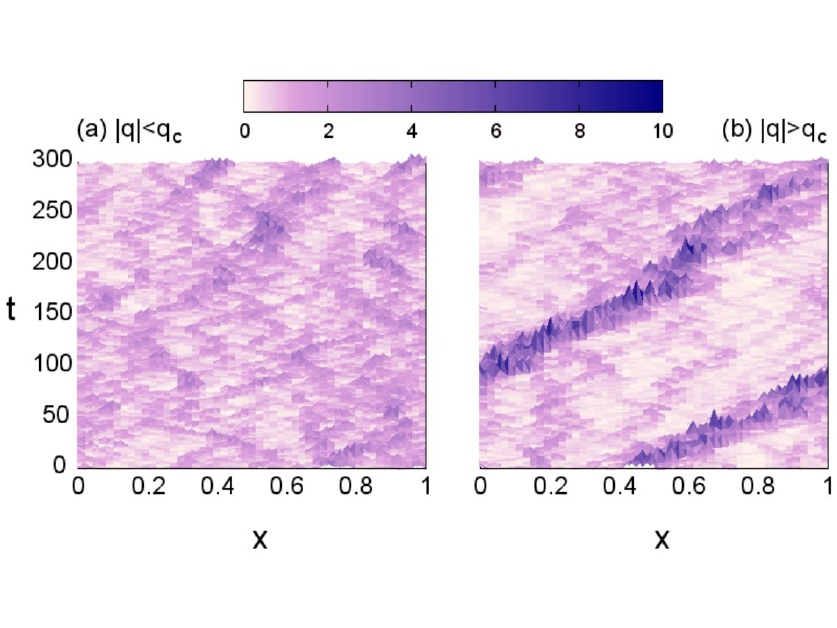}}
\vspace{-1.0cm}
\caption{ (Color online) Typical evolution of the energy field for different current fluctuations in the 1D KMP model on a ring.
(a) Small current fluctuations result from weakly-correlated local events. 
(b) However, for $|q|>q_c$ the system facilitates this unlikely deviation by forming a traveling wave.
}
\label{evolution}
\end{figure}

In this paper we report compelling evidences of this phenomenon in a paradigmatic model of transport 
in one dimension (1D), where we study fluctuations of the time-averaged current.
We find that small current fluctuations result indeed from the sum of weakly-correlated local random events in the density field, 
thus giving rise to Gaussian statistics as dictated by the central limit theorem, see Fig. \ref{evolution}.a. However, for 
large enough currents, the system self-organizes into a coherent traveling wave which facilitates this rare event by accumulating energy in a localized packet, see Fig. \ref{evolution}.b, with a critical current $q_c$ separating both regimes. 
This phenomenon, predicted by HFT \cite{Bertini3,BD2}, is most striking for this model as it happens in 
an isolated equilibrium system in the absence of any external field, breaking spontaneously
a symmetry in 1D. This is an example of the general observation that symmetry-breaking instabilities 
forbidden in equilibrium steady states can however happen at the fluctuating level or in nonequilibrium 
settings \cite{Jona2}. Such instabilities may help explaining puzzling 
asymmetries in nature \cite{Jona2}, from the dominance of left-handed chiral molecules in biology to the matter-antimatter asymmetry in cosmology.

Our model system is the paradigmatic 1D Kipnis-Marchioro-Presutti (KMP) model of transport on a ring \cite{kmp}.
This is a general model of transport which represents at a coarse-grained scale the physics of many quasi-1D systems
of theoretical and technological interest characterized by a single locally-conserved field which diffuses across space.
In this sense our results are of great generality and may have important implications in actual experiments. 
Moreover, this model acts as a benchmark to test theoretical advances in nonequilibrium physics \cite{kmp,Pablo,IFR,Bertini3,Bertini4}.
The model is defined on a 1D lattice of $N$ sites with periodic boundary conditions. Each site $i\in[1,N]$ is characterized by an energy $\rho_i\ge 0$, 
and models an oscillator which is mechanically uncoupled from its nearest neighbors but interacts stochastically with them via a random energy 
redistribution process which conserves total energy $\rho_0\equiv \sum_{i=1}^N \rho_i$.
\begin{figure}[t]
\vspace{-0.6cm}
\centerline{
\includegraphics[width=8cm]{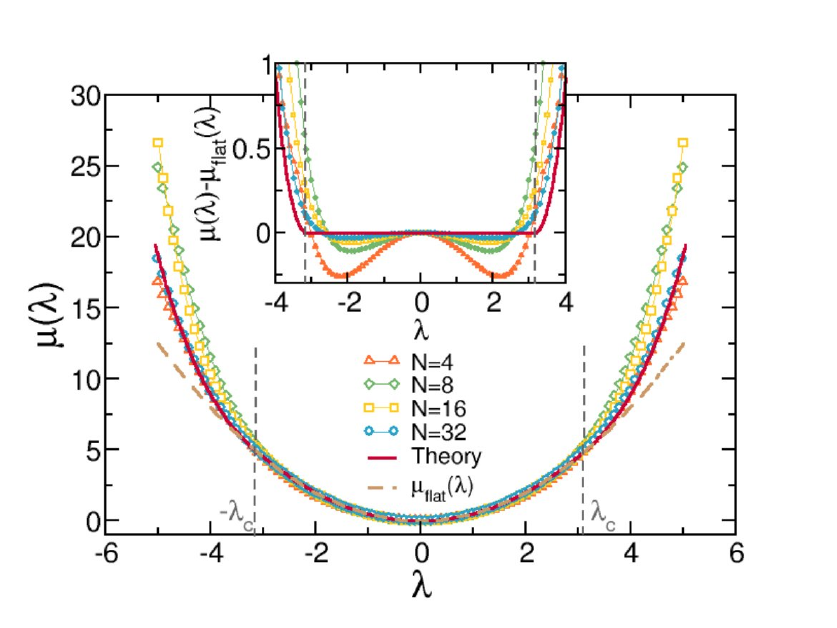}}
\vspace{-0.2cm}
\caption{ (Color online) Main: Measured $\mu(\lambda)$ for the 1D KMP model on the ring and increasing values of $N$, together with
the HFT prediction and the Gaussian approximation. Inset: $\mu(\lambda)-\mu_{\text{flat}}(\lambda)$ for the same $N$. Data converge 
to the HFT prediction as $N$ increases.}
\label{mu}
\end{figure}
We are interested in the statistics of the total current $q$ flowing through the system, averaged over a long diffusive time $\tau$. For $\tau \to \infty$ this time average converges toward the ensemble average $\la q\ra$, which is of course zero because the system is isolated and in equilibrium. However, for long but finite $\tau$ we may still observe fluctuations $q\ne \la q\ra$, and their probability $\text{P}_{\tau}(q)$ obeys a large deviation principle in this limit \cite{LD}, $\text{P}_{\tau}(q)\sim \exp[+\tau N G(q)]$. This means that the probability of observing a current fluctuation decays exponentially as both $\tau$ and $N$ increase, at a rate given by the current LDF $G(q) \le 0$, with $G(\la q\ra)=0$. For KMP model 
a singularity has been shown to exist in $G(q)$ \cite{Bertini3} whose details we uncover here.

In order to study in depth current statistics, we performed extensive simulations of the 1D KMP model with $\rho_0=1$ using an advanced Monte Carlo method which allows to explore the tails of the current LDF \cite{sim,sim2}. This method implies a modification of the stochastic dynamics so that the rare events responsible of a large current fluctuation are no longer rare, and requires the simulation of multiple \emph{clones} of the system \cite{sim}. In this work we used $M=10^4$ clones.  The method yields the Legendre transform of the current LDF, $\mu(\lambda)=\max_q[G(q) + \lambda q]$, with $\lambda$ a parameter conjugated to the current, and Fig. \ref{mu} shows simulation results for $\mu(\lambda)$ and increasing values of $N$. As shown below, HFT predicts Gaussian current statistics for $|q|< q_c=\pi$ --see eq. (\ref{qcrit}), corresponding to quadratic behavior in $\mu(\lambda)= \mu_{\text{flat}}(\lambda)=\lambda^2/2$ up to a critical $|\lambda_c|=\pi$. This is fully confirmed in Fig. \ref{mu} as $N$ increases, 
meaning that small and intermediate current fluctuations have their origin in the superposition 
of weakly-correlated local events, giving rise to Gaussian statistics as dictated by the central limit theorem. However, for fluctuations above the critical threshold, $|\lambda|>\lambda_c$, deviations from this simple quadratic form are apparent, signalling the onset of a phase transition.
In fact, as $N$ increases a clear convergence toward the HFT prediction is observed, with very good results already for $N=32$. Strong finite size effects associated with the finite population of clones $M$ prevent us from reaching larger system sizes \cite{Pablo2}, but $N=32$ 
is already \emph{close} enough to the asymptotic hydrodynamic behavior. Still, small corrections to the HFT predictions are 
observed which quickly decrease with $N$, see inset to Fig. \ref{mu}.

\begin{figure}[t]
\vspace{-0.6cm}
\centerline{
\includegraphics[width=8cm]{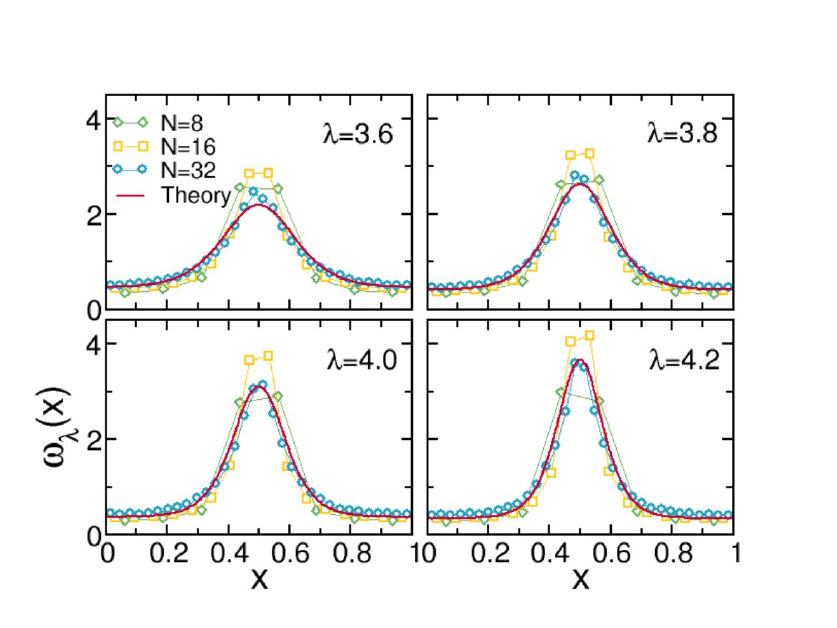}}
\vspace{-0.4cm}
\centerline{
\includegraphics[width=8cm]{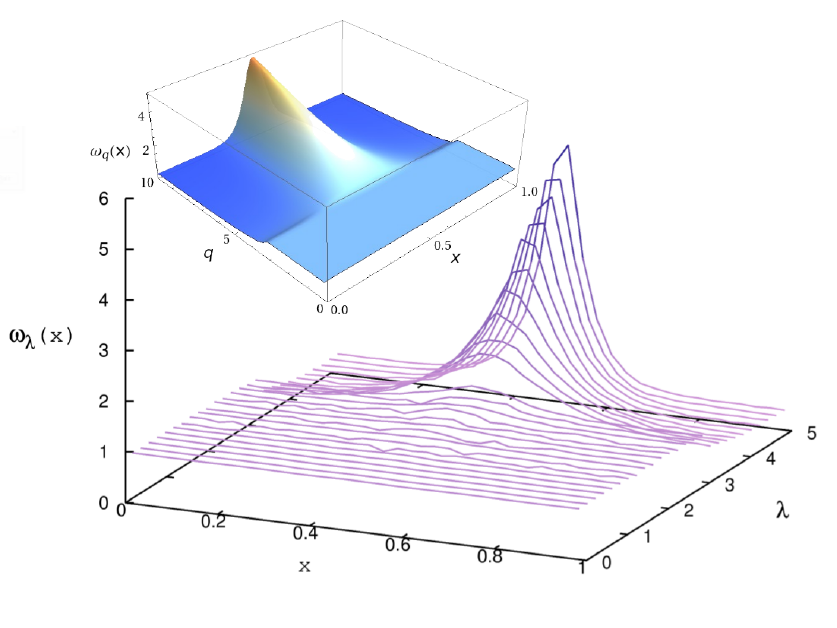}}
\vspace{-0.4cm}
\caption{ (Color online) Top: Supercritical profiles for different $\lambda$ and varying $N$, and HFT predictions.
Bottom: Measured profiles as a function of $\lambda$ for $N=32$. 
Inset: HFT prediction for the optimal profile $\omega_q(x)$. 
Profiles are flat up to the critical current, beyond which a nonlinear wave pattern develops.
}
\label{profsim}
\end{figure}

\begin{figure}[t]
\vspace{-0.6cm}
\centerline{
\includegraphics[width=8cm]{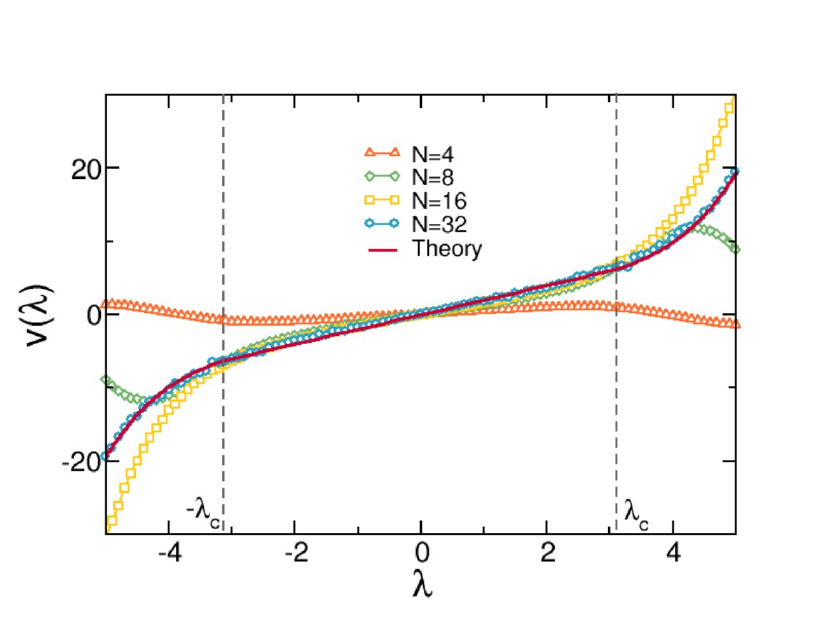}}
\vspace{-0.2cm}
\caption{ (Color online) Velocity measured as a function of $\lambda$ for increasing $N$, and HFT result.}
\label{velo}
\end{figure}

The phase transition is most evident at the configurational level, so we measured the average energy profile associated to a given current fluctuation \cite{sim2}, see 
Fig. \ref{profsim}. 
Due to the system periodicity, and in order not to blur away the possible structure, we performed profile
averages around the instantaneous center of mass. For that, we consider the system as a 1D ring embeded 
in two-dimensional space, and compute the angular position of the center of mass, shifting it to the origin before averaging.
Notice that this procedure yields a spurious weak structure in the subcritical region, equivalent to averaging random profiles 
around their (random) center of mass. Such spurious profile is of course independent of $q$, and can be easily substracted. 
On the other hand, supercritical profiles exhibit a much pronounced structure resulting from the appearance of a traveling wave, 
see Fig. \ref{evolution}.b. Top panel in Fig. \ref{profsim} shows the measured profile $\omega_{\lambda}(x)$ for different 
$\lambda>\lambda_c$ and varying $N$. Again, fast convergence toward the HFT result is observed, 
with excellent agreement for $N=32$ in all cases. Bottom panel in Fig. \ref{profsim} shows the measured 
profiles for $N=32$ and different $\lambda$, which closely resembles the HFT scenario, see inset.
We also measured the average velocity associated to a given current fluctuation by fitting the motion of the center of mass 
during small time intervals $\Delta t$ to a ballistic law, $x_{\text{CM}}(t+\Delta t) - x_{\text{CM}}(t) = v \Delta t$, see e.g. 
Fig. \ref{evolution}.b, and making statistics for the measured velocity. Fig. \ref{velo} shows the mean velocity for $\Delta t=100$ Monte Carlo steps as a function 
of $\lambda$ for increasing values of $N$, and the agreement with HFT is again very good already for $N=32$ (other values of $\Delta t$ 
yield equally good results). Notice that for subcritical current fluctuations the velocity is simply proportional to the current, while 
above the critical line the relation becomes nonlinear.

To understand this behavior, note that the KMP model belongs to a large class of diffusive systems which evolve in time according to a rescaled continuity equation
\be
\partial_t \rho = \partial_x \Big(D[\rho]\partial_x \rho + \xi \Big) \, .
\label{langevin}
\ee
Here $\rho(x,t)$ is the density field, with $x\in[0,1]$, $j(x,t)\equiv -\left(D[\rho]\partial_x \rho + \xi\right)$ is the fluctuating current,
and $D[\rho]$ is the diffusivity (a functional of the density profile in general). The (conserved) noise term $\xi(x,t)$, which accounts for 
microscopic random fluctuations at the mesoscopic level, is Gaussian and white with $\la \xi(x,t)\ra=0$ and 
$\la \xi(x,t)\xi(x',t')\ra=N^{-1} \sigma[\rho] \delta(x-x') \delta(t-t')$, being $\sigma[\rho]$ the mobility functional
and $N$ the system size. In particular, for the KMP model $D[\rho]=1/2$ and $\sigma[\rho]=\rho^2$ \cite{kmp}, and we focus here on 
periodic boundary conditions, 
$\rho(0,t)=\rho(1,t)$ and $j(0,t)=j(1,t)$, so 
the total \emph{mass} in the system is conserved, 
$\rho_0=\int_0^1 \rho(x,t) dx$.
The probability of observing a particular history $\{\rho(x,t),j(x,t)\}_0^{\tau}$ of duration $\tau$
for the density and current fields can be written as a path integral over all possible noise realizations, 
$\{\xi(x,t)\}_0^{\tau}$, weighted by its Gaussian measure, and restricted to those realizations compatible with eq. 
(\ref{langevin}) at every point in space and time. This results in 
$\text{P}(\{\rho,j\}_0^{\tau}) \sim \exp\{+N\, {\cal I}_{\tau}[\rho,j]\}$,
with a rate functional defined by the familiar formula \cite{Bertini,Bertini2,Bertini3,Jona}
\be
{\cal I}_{\tau}[\rho,j] = - \int_0^{\tau} dt \int_0^1 dx \frac{\displaystyle \Big(j+D[\rho]\partial_x\rho \Big)^2}{\displaystyle 2\sigma[\rho]} \, .
\label{HFT2}
\ee
with $\rho(x,t)$ and $j(x,t)$ coupled via the continuity equation, $\partial_t \rho + \partial_x j=0$.
Eq. (\ref{HFT2}) expresses the locally Gaussian nature of current fluctuations around its average behavior, given by Fourier's law.
We are interested in the fluctuations of the time-averaged current $q=\tau^{-1} \int_0^{\tau} dt \int_0^1 dx j(x,t)$.
The probability of observing a given $q$ can be in turn obtained from the path integral of $\text{P}(\{\rho,j\}_0^{\tau})$ restricted to histories
$\{\rho,j\}_0^{\tau}$ consistent with that value of $q$. This probability scales as
$\text{P}_{\tau}(q)\sim \exp [+\tau N G(q)]$, 
and the current LDF $G(q)$ is related to ${\cal I}_{\tau}[\rho,j]$ via a simple
saddle-point calculation in the long time limit, $G(q)=\tau^{-1} \max_{\rho,j}{\cal I}_{\tau}[\rho,j]$, such that the \emph{optimal} profiles $\rho_q(x,t)$ 
and $j_q(x,t)$ solution of this variational problem are compatible with the constraints on $\rho_0$ and $q$
and are related via the continuity equation. These optimal profiles can be interpreted as the 
ones the system adopts to facilitate a given current fluctuation.

Small deviations of the empirical current away from its ensemble average $\la q\ra=0$
typically result from weakly-correlated local fluctuations. 
The average density profile associated to these small fluctuations hence corresponds still to the flat, stationary one, 
$\rho_q(x,t)=\rho_0$. In this case, the optimal current profile is just $j_q(x,t)=q$ and the current LDF is simply quadratic, 
$G_{\text{flat}}(q)=-q^2/2\sigma(\rho_0)$ \cite{BD2}, resulting in Gaussian current statistics as confirmed is simulations, see Fig. \ref{mu}.
A natural question thus concerns the stability of this flat profile against small perturbations. 
Bodineau and Derrida have shown \cite{BD2} that the flat profile indeed becomes unstable, in the sense that 
$G(q)$ increases by adding a small time-dependent periodic perturbation to the otherwise constant profiles, whenever 
$8 \pi^2 D(\rho_0)^2 \sigma(\rho_0) - q^2\sigma''(\rho_0) < 0$, where $\sigma''$ denotes second derivative. This defines a critical current 
\be
|q_c|=2\pi D(\rho_0) \sqrt{2\sigma(\rho_0)/\sigma''(\rho_0)}
\label{qcrit}
\ee
for the instability to kick in. When this happens, the form of the associated relevant perturbation suggests that current 
fluctuations in this regime are sustained by a traveling wave pattern moving at constant velocity $v$ \cite{BD2}, as observed in simulations. We hence write
$\rho_q(x,t)=\omega_q(x-vt)$, which results in 
$j_q(x,t)=q-v\rho_0+v\omega_q(x-vt)$ via the continuity equation. The variational problem for $G(q)$ can be now written as
\be
G(q) = -\min_{\omega_q(x), v} \int_0^1 \frac{\displaystyle \big[q-v\rho_0+v\omega_q(x) \big]^2 + \omega'_q(x)^2 D[\omega_q]^2}
{\displaystyle 2\sigma[\omega_q] }\, dx
\label{ldfwave}
\ee
resulting in the following differential equation for the shape of the optimal traveling wave
\be
\big[q-v\rho_0+v\omega_q(x) \big]^2 - \omega'_q(x)^2 D[\omega_q]^2 = 2\sigma[\omega_q] \Big\{ C_1 + C_2 \omega_q(x) \Big\} \, .
\label{weq}
\ee
This equation yields a $\omega_q(x)$ which is generically a symmetric function with a single minimum $\omega_1=\omega(x_1)$ and
maximum $\omega_2=\omega(x_2)$ such that $|x_2-x_1|=1/2$. The constants $C_1$ and $C_2$ can be then related to these extrema, 
which in turn are fixed by the constraints on the total mass of the system and the distance between extrema,
\be
\frac{\rho_0}{2}=\int_{\omega_1}^{\omega_2} \frac{\omega D(\omega)}{Z_v(\omega)} d\omega  \quad ; \quad 
\frac{1}{2}=\int_{\omega_1}^{\omega_2} \frac{D(\omega)}{Z_v(\omega)} d\omega \, ,
\label{constraints}
\ee
where $Z_v(\omega)=[(q-v\rho_0+v\omega)^2 - 2\sigma(\omega)(C_1+C_2\omega)]^{1/2}$.
The optimal wave velocity is given implicitely by $v=-q \nu_1^{(v)}/\nu_2^{(v)}$ \cite{BD2}, with
\be
\nu_n^{(v)}\equiv \int_{\omega_1}^{\omega_2} \frac{\displaystyle D(\omega) \big(\omega-\rho_0\big)^n}
{\displaystyle \sigma(\omega) Z_v(\omega)}\, d\omega
\label{veq}
\ee
In this way, for constant $\rho_0$ and $q$, we use eqs. (\ref{constraints})-(\ref{veq}) to compute the profile extrema and
its velocity, and use this information to solve eq. (\ref{weq}) for $\omega_q(x)$. 
The resulting predictions are fully confirmed in simulations, see above.

Our results unambiguously show that an isolated diffusive system exhibits a phase transition at the fluctuation level. This phenomenon, captured by hydrodynamic fluctuation theory, is most surprising as it happens in an equilibrium system in the absence of external fields, breaking spontaneously a symmetry in 1D. This illustrates the idea that critical phenomena not allowed in equilibrium steady states may however arise in their fluctuating behavior or under nonequilibrium conditions \cite{Jona2}. Remarkably, similar instabilities have been described in quantum systems \cite{quantum}. Our results strongly support that the phase transition is continuous as conjectured in \cite{BD2}, excluding the possibility of a first-order scenario, and suggest that a traveling wave is in fact the most favorable time-dependent profile in the supercritical regime. This observation may greatly simplify general time-dependent calculations, but the question remains to  whether this is the whole story or other, more complex solutions may play a dominant role for even larger fluctuations. In any case, it seems clear that rare events call in general for coherent, self-organized patterns in order to be sustained \cite{rare}.

Financial support from Spanish MICINN project FIS2009-08451, University of Granada, and Junta de Andaluc\'{\i}a projects P07-FQM02725 and P09-FQM4682 is acknowledged.

\end{document}